\documentclass[aps,prb,twocolumn,reprint,superscriptaddress,groupeaddress, showpacs, letterpaper, citeautoscript]{revtex4-1}
	
\usepackage{amsmath}
\usepackage{amssymb}

\usepackage{CJK}
\usepackage{graphicx}

\begin{document}

\begin{abstract}
Materials exhibiting reversible resistive switching in electrical fields are highly demanded for functional elements in oxide electronics. In particular, multilevel switching effects allow for advanced applications like neuromorphic circuits. Here we report on a structurally driven switching mechanism involving the so-called ``dead'' layers of perovskite manganite surfaces. Forming a tunnel barrier whose thickness can be changed in monolayer steps by electrical fields, the switching effect exhibits well-defined and robust resistive states.
\end{abstract}

\title{Layer-by-layer resistive switching: multi-state functionality due to electric-field-induced healing of ``dead'' layers}

\author{Jon-Olaf Krisponeit}
\email[]{krisponeit@ifp.uni-bremen.de}
\affiliation{Institute of Solid State Physics, University of Bremen, Otto-Hahn-Allee 1, 28359 Bremen, Germany}
\affiliation{I. Physikalisches Institut, Georg-August-Universit{\"a}t G{\"o}ttingen,  37077 G{\"o}ttingen,  Germany}
\affiliation{MAPEX Institute for Materials and Processes, University of Bremen,  28359 Bremen, Germany}
\author{Bernd Damaschke}
\affiliation{I. Physikalisches Institut, Georg-August-Universit{\"a}t G{\"o}ttingen,  37077 G{\"o}ttingen,  Germany}
\author{Vasily Moshnyaga}
\affiliation{I. Physikalisches Institut, Georg-August-Universit{\"a}t G{\"o}ttingen,  37077 G{\"o}ttingen,  Germany}
\author{Konrad Samwer}
\affiliation{I. Physikalisches Institut, Georg-August-Universit{\"a}t G{\"o}ttingen,  37077 G{\"o}ttingen,  Germany}


\maketitle

Resistance switching effects driven by electric fields are raising hope for novel multi-state memory applications and, mimicking biological synapses, for advances in neuromorphic computing.\cite{Yang2012, Waser2007} Functional oxides meet their demand for non-binary resistance changes covering several orders of magnitude by providing mechanisms as diverse as displacive effects\cite{Yang2008}, ferroelectric domain switching\cite{Chanthbouala2012} and the tuning of Schottky barriers \cite{Bessonov2015}. These approaches rely on the accurate fabrication of oxide heterostructures. Avoiding such complexity, we demonstrate here that multi-state functionality can be obtained already in a single material. Perovskite manganites are known to exhibit so-called `dead' layers\cite{Jin2016,Borges2001,Huijben2008,Sun1999,Kim2011} forming at interfaces and the free surface. We show that such dead layers can indeed play a most active role: Being switchable in a layer-by-layer fashion between two intrinsically well-defined resistance states, they might pave the way to novel  concepts for robust switching devices as step-wise tunable tunnel barriers.

The resistive switching behavior at the free surface of an epitaxial manganite thin film grown on MgO(100) has been analyzed by conductive atomic force microscopy (C-AFM). The film with nominal composition $\mathrm{La_{0.85}Sr_{0.15}MnO_3}$ was grown by means of the metalorganic aerosol deposition (MAD) technique \cite{Moshnyaga1999} on a polished MgO substrate. Acetylacetonates of La-, Sr- and Mn were used as precursors, solved in dimethylforamide and sprayed onto the heated substrate ($\sim900\,\mathrm{^\circ°C}$ in ambient atmosphere) by compressed air. The molar ratio of precursors in the solution was chosen to obtain the necessary film stoichiometry with Sr-doping level, $\mathrm{x}=0.15$, as was empirically calibrated by means of X-ray diffraction (XRD) and energy dispersive X-ray microanalysis (EDX). According to the XRD, the film shows an out-of-plane epitaxy on MgO with c-axis lattice constant $c=0.3899\,\mathrm{nm}$, which is very close to the bulk LSMO value for $x=0.15-0.17$ \cite{Urushibara1995}, indicating a strain-free state of the film. A thickness of $d\approx60\,\mathrm{nm}$ was estimated from the used precursor amount, calibrated from similar LSMO films by using small-angle X-Ray reflectivity. Resistance measurements reveal an insulator-metal transition at $T_\mathrm{MI}=290\,\mathrm{K}$ and a metal-like behavior at low temperatures with high residual resistance possibly due to a phase separation as the chosen composition is close to the MI phase boundary in the phase diagram of LSMO\cite{Urushibara1995}. 

\begin{figure}[htbp]
\includegraphics[keepaspectratio,width=\columnwidth]{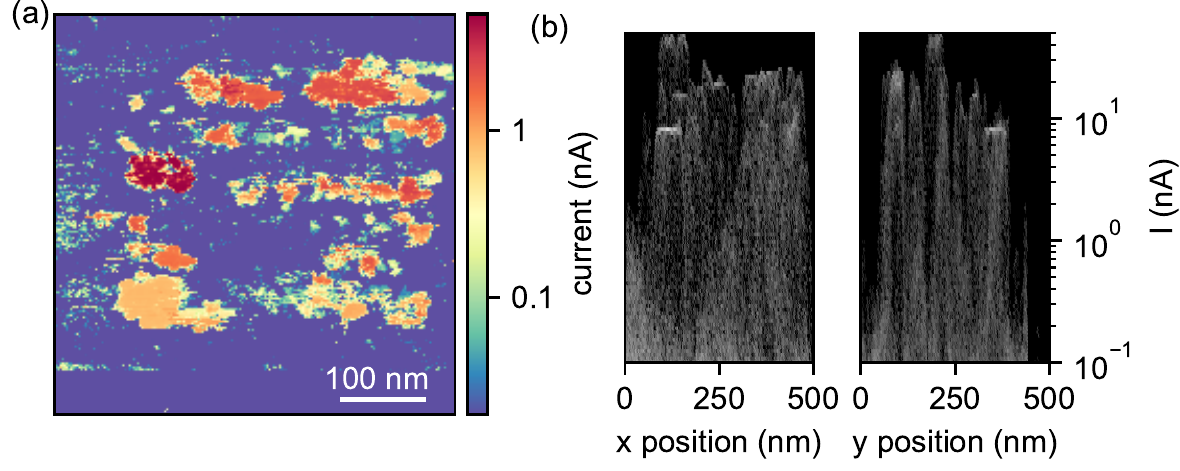}
\caption{Metallic regions created by electric pulses at the initially insulating surface. (a) Current map recorded at $\mathrm{0.1\,V}$ showing the spatial distribution of conductance at the surface. (b) Distribution of current values with respect to their $x$ and $y$ position, visualizing regions of homogeneous conductance as horizontal lines.}\label{fig:1}
\end{figure}

All switching experiments were conducted with an Omicron VT-AFM under ultra-high vacuum conditions (base pressure $<$3x10\textsuperscript{-10} mbar). Commercial probes from MikroMasch with a spring constant of 0.15 N\slash m and a Pt coating were used for C-AFM measurements. Note that due to the coating the tip radius is enlarged, leading to reduced topographic resolution in comparison with standard AFM probes. Measurements were recorded at vertical force setpoints $\leq5\,\mathrm{nN}$, all current maps shown were recorded at a positive tip bias of $0.1\,\mathrm{V}$. Scanning the sample surface with a conductively coated tip, C-AFM enables local switching experiments on a nanometer scale and the simultaneous recording of topography and current maps. It has hence proven indispensable for microscopic investigations on memristive effects \cite{Chen2006,Krisponeit2010,Moreno2010,Munstermann2010,Kalkert2011,Krisponeit2013}. As reported in our previous studies on comparable epitaxial \cite{Krisponeit2010,Krisponeit2013} and nanocolumnar \cite{Kalkert2011} manganite thin films, the LSMO surface exhibits an initial state of a very high resistance (HRS), which can be switched to a low resistive state (LRS) by application of a positive electric pulse above a critical threshold voltage to the tip. The HRS can be restored again by a voltage pulse of the reversed (negative) polarity. A current map of conductive surface regions created at the initially insulating surface by positive voltage pulses is given in Fig.~\ref{fig:1}(a). The  domain distribution was obtained by 64 positive voltage pulses, applied in a regular square pattern with increasing pulse voltages, ranging from $+3\,\mathrm{V}$ to $+6.5\,\mathrm{V}$, from bottom to top and pulse durations increasing from $1\,\mathrm{ms}$ to $3\,\mathrm{s}$ from left to right, see also Ref.~\onlinecite{Krisponeit2010} for comparison.The irregular lateral shape of conducting domains originates from their cascade-like growth dynamics dominated by domain wall pinning at defects and the surface morphology \cite{Krisponeit2013}.

Interestingly, beyond previously reported manganite switching characteristics, we observe here the formation of apparently discrete conductance levels on a microscopic scale: Although the C-AFM tip was at rest during the voltage pulses, the regions of homogeneous conductance were extended to significantly larger dimensions (up to $100\,\mathrm{nm}$) than the actual electrical contact (few nm in diameter). This observation not only excludes a filamentary process for the given case, but the existence of sharp domain boundaries furthermore also implies an abrupt transition proceeding through the material instead of a gradual change. For further analysis, the frequency of the measured conductance values was visualized depending on the $x$ and $y$ positions of the tip (Fig.~\ref{fig:1}(b)). In these intensity maps, the discrete conductance values can be clearly discerned as horizontal lines. These values scatter seemingly arbitrarily, predominantly in the upper part of the instrumental current range ($\mathrm{50\,nA}$) and no correlation between current and sample topography was found.

\begin{figure}[htb!!!]
\includegraphics[keepaspectratio,width=\columnwidth]{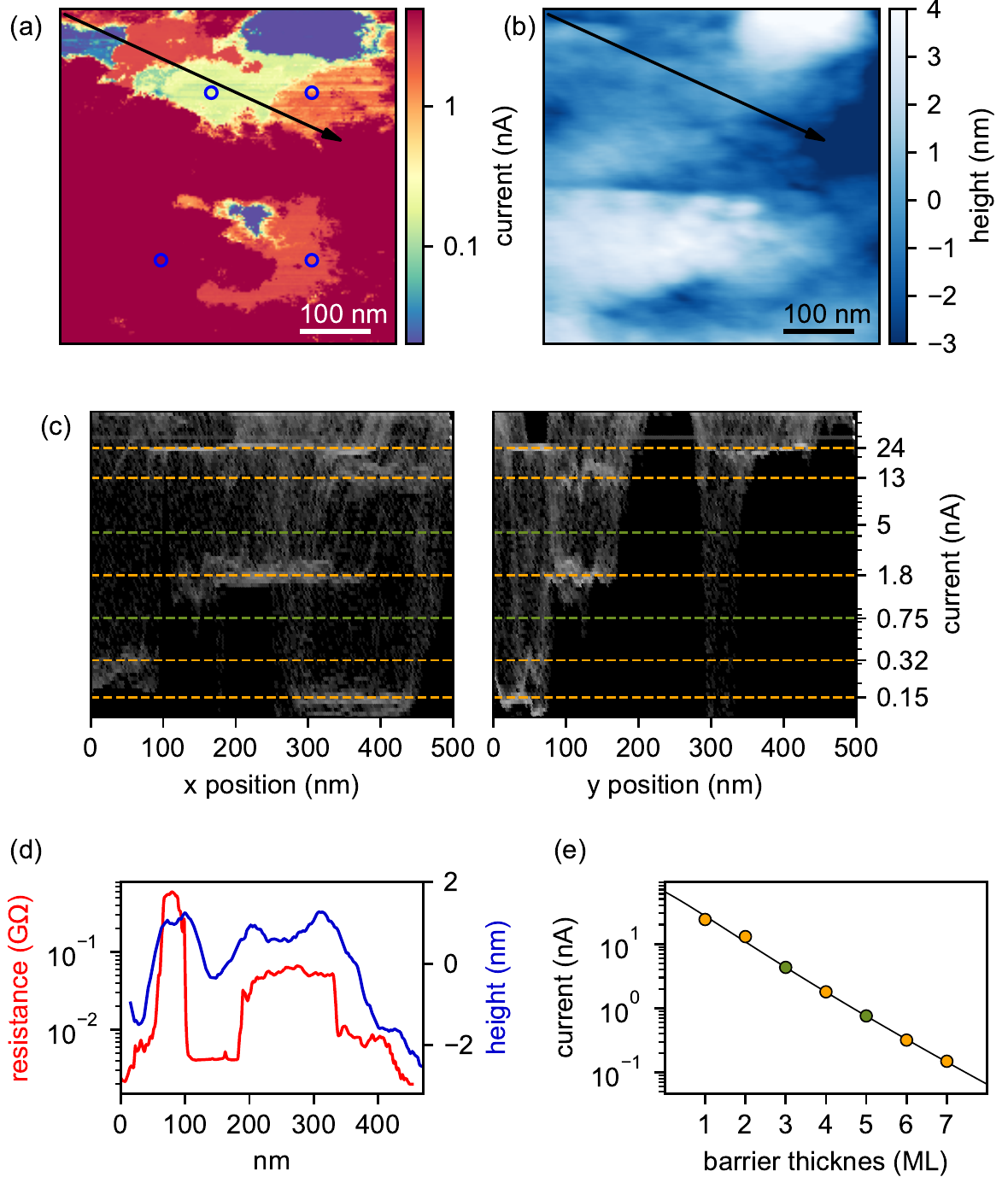}
\caption{Conductance and topography regions switched from LRS to HRS. (a) Current map recorded at U = 0.1 V. After switching to LRS (red environment), negative voltage pulses have been applied at the locations indicated by circles. (b) Corresponding topography image. The correlation between contact resistance (red) and height (blue) is illustrated by profiles along the indicated path (d). (c) Distribution of current values occurring in dependence of the x and y locations as intensity map. Identified logarithmically spaced current values are shown in orange, missing levels are indicated in green. (e) Current in dependence of the ascribed tunnel barrier thickness with fit curve according to MIM model.}\label{fig:2}
\end{figure}

For investigations on the reversed switching direction from LRS to HRS, another sample region has been set to the metallic surface state by several fast scans ($1000\,\mathrm{nm}$) under a positive electrical tip bias of $6\,\mathrm{V}$. By using this technique instead of prolonged pulses with the cantilever resting at a single site, we could avoid thermal damaging of the surface impending from local Joule heating in the LRS. As confirmed via simultaneously recorded current maps, the low resistive state has been created successfully on the entire scan area. Next, voltage pulses of reversed (negative) polarity have been applied at four cantilever positions, whereupon two extended areas of increased resistivity have emerged: one around the lower right pulse position and one including both upper pulse positions in Fig.~\ref{fig:2}(a). No successful switching was observed at the lower left pulse position. As is obvious from the current map, both areas comprise subregions of homogeneous conductivity. Furthermore, though disjunct, each area contains one subregion belonging to the same current level ($\approx 25\,\mathrm{nA}$).

Comparison with the topography data (Fig.~\ref{fig:2}(b)), as further illustrated by line profiles of height and resistance (Fig.~\ref{fig:2}(d)), reveals a remarkable correlation between the sample height and the magnitude of the discrete resistance levels. Regions of lower (higher) resistance are found primarily in  shallow (protruding) regions, respectively. Taking into account that the vertical length scale spans only few lattice constants, we hence claim that the discrete current values stem from insulating barriers with thicknesses of  integer numbers of atomic layers. Contrasting the initial switching process discussed above, the intensity maps (Fig.~\ref{fig:2}(c)) now suggest a non-random, logarithmic spacing of the possible current values, indicating the insulating barrier to act as a tunnel barrier. We therefore ascribe the HRS levels to the existence of so-called ``dead" layers. These regions, where magnetic and electric properties deviate from the bulk, are well-documented to occur on manganite films at interfaces as well as at their free surfaces and result in a tunnel barrier. Reports agree in estimating typical dead layer thicknesses of about 3 to 4 nm for LSMO films under different strain influences arising from a variety of chosen substrates \cite{Borges2001, Huijben2008, Jin2016, Kim2011, Sun1999}. While the exact physical origin for the formation of dead layers is still under discussion, the symmetry break at interfaces and variations in the oxygen content have been identified as key factors. Recent investigations on LSMO elucidated the interfacial loss of inversion symmetry to lift the degeneracy of the $3d\!-\!e_g$ orbitals, resulting in a preferential occupation of the out-of-plane $3z^2-r^2$ orbitals and the emergence of a C-type antiferromagnetic layer \cite{Pesquera2012, Valencia2014}.

\begin{figure}[htbp]
\centering
\includegraphics[keepaspectratio,width=\columnwidth]{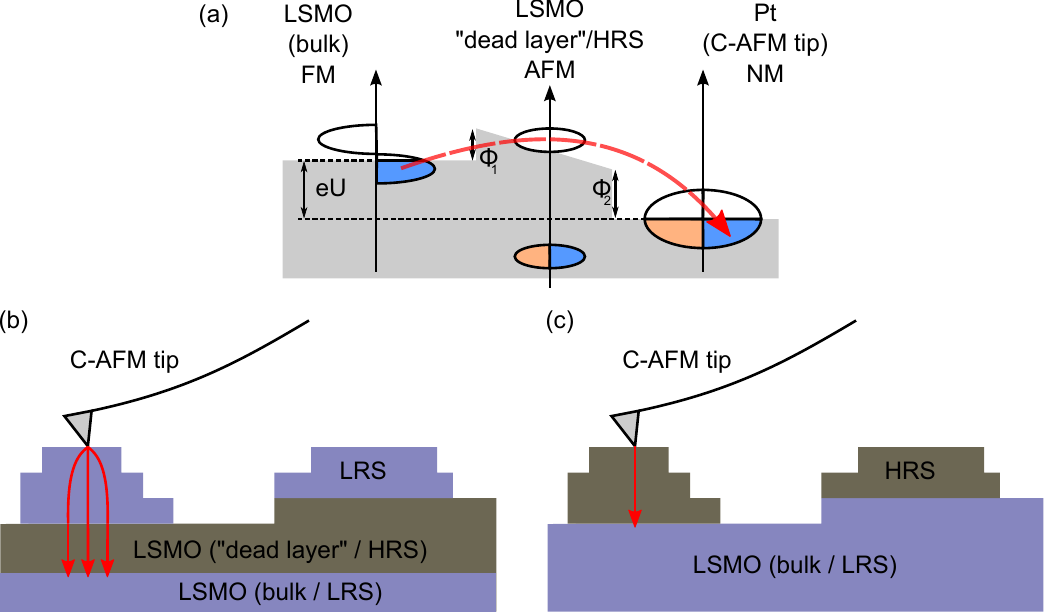}
\caption{Schematic band picture of the intrinsic tunnel barrier and geometrical conctact model of the regions switched in C-AFM. (a) The ferromagnetic metallic bulk of LSMO and the Pt-coated AFM tip are separated by an insulating tunnel barrier. (b) Switching to LRS creates metallic regions at the surface, which act as broadened electrode to the underlying barrier. (c) Applying negative voltage pulses at a fully LRS-switched sample creates insulating layers of integer monolayer thicknesses at the surface, giving rise of the logarithmically step-wise scaling of the current values.}\label{fig:3}
\end{figure}

Our results imply that such a dead layer can indeed be ``healed'' by application of positive voltage pulses, and the ``dead'' HRS state is forming again under reversed polarity both proceeding in a layer-by-layer fashion. A simplified sketch is shown in Fig.~\ref{fig:3}(a). As an ideal half-metal, the LSMO bulk exhibits an exchange splitting of the conduction band formed by the $3d\!-\!e_g$ states. At the surface, we assume a dead layer with strong antiferromagnetic correlations and a gap opening between an upper and lower Hubbard band. Due to this gap, an intrinsic tunnel barrier is formed between bulk LSMO and the top electrode, the Pt-coated AFM tip. The width of this barrier and therewith the tunneling conductance can be modified by electric fields. Note that in this scenario only two well-defined intrinsic resistive states are involved and the observed multitude of ``device'' resistance levels originates solely from the varying number of layers forming the insulating barrier at the surface. This picture contains strong conceptual similarities with switchable ferroelectric and multiferroic tunnel junctions \cite{Pantel2012,Chanthbouala2012, Gajek2007}. The combination of the LSMO bulk, which is also often utilized as a half-metallic lead material in the latter case, and the functional surface layer could be regarded as an ``intrinsic heterostructure'' showing related physical phenomena. However, requiring a precise description of the complex interplay with the $t_{2g}$ states, the $O\, 2p$ band and structural correlations, the derivation of a detailed microscopic model is obviously beyond the scope of this study. Interestingly, Shen et al. very recently proposed a possible magneto-structural mechanism for the switching of dead layers---in that case resulting in a suppression of spin-polarized transport in half-metallic LCMO \cite{Shen2016}.

In their recent work, Norpoth et al. \cite{Norpoth2014} provided evidence that in the case of $\mathrm{Pr_{0.7}Ca_{0.3}MnO_3}$ resistive switching processes can actually involve a combination of structural transitions and oxygen vacancy migration. In our case, the results indicate the structural\slash electronic transition of LSMO dead layers to act as the fundamental switching step. This is clearly a distinct mechanism, which acts independently at the surface of the sample. Diffusive switching processes affecting larger sample fractions can yet be stimulated by, e.g., high current densities and Joule heating due to the switched surface. For future applications, however, the opportunity to exploit an abrupt transition between two well-defined states might promise extraordinary device robustness on small length scales.
 
Let us now review the phenomenological model of surface dead layers and consider the discrete current values to reflect tunnel barriers of integer monolayer thickness. Therefore, the resulting metal-insulator-metal (MIM) architecture was treated within tunneling theory. For a proof of the model, we tentatively assigned an increasing layer thickness of perovskite monolayers to the current values in descending order. Five current values could be reliably indentified in the corresponding intensity maps (Fig.~\ref{fig:2}(d)). In addition, the maps suggest only logarithmically spaced current values to be allowed. We believe that two of these possible current levels have not been realized in the experiment, which have therefore been included as interpolated values. As illustrated in Fig.~\ref{fig:2}(e), the current values decrease approximately exponentially with the assumed layer thickness, confirming the existence of a tunnel barrier. One has to keep in mind, however, that no direct determination of the barrier thickness is possible from the experiment and that the minimal barrier thickness of one ML chosen here is arbitrary. The same exponential behavior would be observed for an offset in the ascribed layer thickness. Barriers of even higher thickness are difficult to be observed in our experiments, because the tunneling current would lie at or beneath the instruments noise level. Yet, the investigated range of 7 ML ($\approx 2.7\,\mathrm{nm}$) agrees well with the typical thicknesses of dead layers estimated in the aforementioned studies. The height of the tunnel barrier at the hidden interface, $\phi_I$, has been estimated according to the formula derived by John G. Simmons for MIM architectures \cite{Simmons1963a, Simmons1963b}. In our case, a satisfactory fit result (see Fig.~\ref{fig:2}(e)) could be obtained only for the case of high voltages, i.e.\ where $eV>\phi_{II}$, with $\phi_{II}$ being the barrier height between the LSMO surface and the Pt AFM tip (Fig.~\ref{fig:3}(a)). The current is given by: 
\begin{equation}
I=\frac{A\kappa}{\phi_I(z+z_0)^2} \cdot \Big( e^{-\lambda(z+z_0)\phi_I^{3/2}} - b \cdot e^{-\lambda(z+z_0)\phi_I^{3/2}\cdot b}\;\Big).
\end{equation}
Here,  $\kappa=\frac{e^3U^2}{8\pi^2\hbar}$ and $\lambda = \frac{2\beta\sqrt{2m}}{eU\hbar}$, depending only on the applied tip voltage, are known constant, where $\beta=\frac{23}{24}$ for a triangular barrier shape \cite{Simmons1963b} and $b=\sqrt{1+2eU/\phi_I}$. For the effective mass of the charge carriers in LSMO the value estimated by Okuda et al. of $m \approx 2.5 m_0$ (Ref. \cite{Okuda1998}) was used. $U$ is the bias voltage applied to the cantilever tip. Considering the existence of a minimal barrier thickness, an additional offset $z_0$ has been taken into account for $z$. Three physical fitting parameters have been derived: The barrier height $\phi_I$, the minimum barrier thickness $z_0$ and the electrical contact area $A$. The latter two parameters are strongly correlated and cannot be reliably determined from the fitting procedure. Nevertheless, we obtained reasonable values for $A\approx2.0\,\mathrm{nm^2}$ (see Ref. \cite{Krisponeit2010} for an estimation) and $z_0 \approx 1.3\,\mathrm{nm}$, further validating the model. The barrier height has been determined as $\phi_I=46\,\mathrm{meV}$. Due to the uncertainties regarding $A$ and $z_0$, the limited data available, and, first of all, the not precisely known effective mass, we have to consider this value as a rough estimate but trust in $\phi_I$ being below $60\,\mathrm{meV}$.

Being remarkably small, this barrier height between bulk LSMO and the dead layer might be indicative for band bending effects but also for a Fermi level pinning closely below the conduction band inside the dead layer. This behavior would not constitute a contradiction to previous results, as significant Fermi level pinning has been excluded only for bulk LSMO which lacks the tendency for intrinsic electronic phase separation. While the minimal barrier thickness cannot be accurately determined from the given fit and some thickness values might be allowed yet unrealized in the presented experiments, it is yet remarkable that the obtained value of $1.3\,\mathrm{nm}$ is in perfect agreement with the results of Valencia et al. of a 2-4 ML thick very robust C-type antiferromagnetic layer at LSMO interfaces \cite{Valencia2014}.

Finally, we must address the question why logarithmically spaced current values have been observed only for the switching direction from LRS to HRS and not vice versa, where nevertheless patches of homogeneous conductance were found. Indeed, a careful investigation based of the contact geometry provides a phenomenological understanding of this behavior: The initial sample state has to be regarded as sketched in Fig.~\ref{fig:3}(b). The dead layer reaches down to an intrinsical, characteristic depth. It is natural to assume the energetical preference of a flat interface between bulk and dead layer, hence small surface corrugations lead to locally increased barrier thicknesses. Considering the switching from HRS to LRS to be initiated at the tip---where the electric field strength is the strongest---switched metallic regions grow into the sample. Depending on pulse strength, duration and pinning processes at defects and, in particular, the surface morphology, a metallic region of variable size grows atop of the dead layer. Note that in a previous study on compatible manganite films we have demonstrated the importance of domain wall pinning and depinning effects in the evolution of metallic regions as reflected in the occurrence of avalanche-like dynamics \cite{Krisponeit2013}.

As a consequence, a MIM structure evolves with an effectively enlarged top electrode of random size. The observed current values depend on its lateral dimensions as well as on the reduced dead layer thickness remaining underneath. This picture therefore meets the observed randomness and the generally higher current values as observed in the experiment. On the other hand, having switched the entire surface region to the LRS it is most intuitive to assume the switching process to initiate at the topmost layers again proceeding layerwise into the sample (Fig.~\ref{fig:3}(c)). Now, only the C-AFM tip acts as top electrode --- probing the insulating layer thickness with a constant tip-to-sample contact area and hence depicting the logarithmically scaled current values.

To conclude, we have presented evidence for a layer-by-layer proceeding resistive switching. For the underlying physical mechanism, the existence of dead layers acting as tunnel barriers appears to be a crucial factor. The layer-wise appearance further suggests for resistive switching at the LSMO surface an electro\slash magneto-structural transition rather than a diffusive mechanism. From an applicational perspective, the existence of two inherent, well-defined resistivity states and the demonstrated possibility to actively control them on a monolayer scale might trigger novel approaches to resistive switching devices featuring improved device lifetimes due to the robustness against chemical and structural degradation of the structural mechanism.
The new, step-wise switching mode provides a solid multi-state functionality as desired for advances in oxide electronics and neuromorphic computing architectures.

The authors gratefully acknowledge financial support from the DFG via SFB 602 TP A2 and the Leibniz program. JOK has been supported by the Institutional Strategy of the University of Bremen, funded by the German Excellence Initiative. All authours greatly acknowledge fruitful discussion with Christin Kalkert and assistance with the sample preparation and characterization by Francesco Belfiore (University of California, Santa Cruz). JOK would like to thank Jan H{\"o}cker and Jan Ingo Flege (University of Bremen) for helpful discussions.

\end{document}